\documentclass[prb,showpacs,twocolumn,superscriptaddress]{revtex4}
\pdfoutput=1
\usepackage{amsmath,latexsym,natbib,bm,psfrag,color,amsmath,overpic,rotating}
\usepackage{epstopdf}

\begin{document}

\title{ First-principles simulations on the structural and energetic properties
        of domains in PbTiO$_3$/SrTiO$_3$ superlattices}

\author{ Pablo Aguado-Puente}
\affiliation{ Departamento de Ciencias de la Tierra y
              F\'{\i}sica de la Materia Condensada, Universidad de Cantabria,
              Cantabria Campus Internacional,
              Avenida de los Castros s/n, 39005 Santander, Spain}
\author{ Javier Junquera }
\affiliation{ Departamento de Ciencias de la Tierra y
              F\'{\i}sica de la Materia Condensada, Universidad de Cantabria,
              Cantabria Campus Internacional,
              Avenida de los Castros s/n, 39005 Santander, Spain}

\date{\today}

\begin{abstract}
 We report first-principles calculations, within the density functional
 theory, on the structural and energetic properties of 
 180$^\circ$ stripe domains in 
 (PbTiO$_{3}$)$_{n}$/(SrTiO$_{3}$)$_{n}$ superlattices.
 For the explored periodicities ($n$=3 and 6) we find that 
 the polydomain structures compete in energy with the monodomain phases.
 Our results suggest the progressive transition, as a function of $n$,
 from a strong to a weak electrostatic coupling regime between the 
 SrTiO$_{3}$ and PbTiO$_{3}$ layers.
 Structurally, they display continuous
 rotation of polarization connecting 180$^\circ$ domains.
 A large offset between [100] atomic rows across the domain wall and 
 huge strain gradients are observed. 
 The domain wall energy is very isotropic, depending very weakly on the
 stripe orientation.
\end{abstract}

\maketitle

\section{Introduction}

 The growth of superlattices composed of thin layers of ABO$_{3}$ 
 perovskites with different physical properties has become 
 one the the most promising paths to exploit the coupling
 between instabilities in order to engineer
 new functionalities in these 
 heterostructures.~\cite{Dawber-05,Ghosez-06,Zubko-11}
 For a long time the focus was on the electrostatic coupling
 between the layers of the superlattice,~\cite{Neaton-03,Nakhmanson-05} 
 and the interplay with the epitaxial strain.~\cite{Schlom-07}
 More recently, after the discovery of the appearance of a polarization
 from the coupling of two rotational modes in ultra-short period
 PbTiO$_3$/SrTiO$_3$ superlattices,~\cite{Bousquet-08} the interest has evolved 
 to include also the
 interaction between ferroelectric (FE) and antiferrodistortive (AFD) modes in 
 perovskite related layered 
 compounds.~\cite{Aguado-Puente-11, Benedek-11,Rondinelli-11}

 At the same time, polarization domains in ferroelectric thin
 films are being subject of numerous investigations due to the
 recent finding of their intrinsic functional properties.
 One of the most remarkable is the conductivity at domain 
 walls within an otherwise insulating 
 material.~\cite{Seidel-09,Guyonnet-11}
 The great activity in the research of polydomain phases in ferroelectric
 thin films has also demonstrated that the domain structure in
 these systems, with rotation of the polarization~\cite{Catalan-06, Lee-09} 
 or the formation of flux-closure structures
 in thin films,~\cite{Lai-07,Aguado-Puente-08, 
 Prosandeev-10, Ivry-10, Jia-11, Nelson-11}  
 and vortex structures in nano-sized ferroelectrics,~\cite{Naumov-04,Prosandeev-08.2} 
 probably differs significantly from the typically assumed picture 
 of alternating regions where the polarization points along opposite directions
 with sharp domain walls (DW).

 In the particular case of PbTiO$_3$/SrTiO$_3$ 
 interfaces, pioneer works were devoted to the identification of 
 180$^{\circ}$ stripe domains in PbTiO$_{3}$ \emph{thin films} 
 grown on thick SrTiO$_{3}$ (001) substrates.
 The domain structures were characterized
 both in reciprocal space (strong satellites around
 PbTiO$_{3}$ Bragg peaks in synchrotron x-ray scattering 
 measurements),~\cite{Streiffer-02,Stephenson-03,Fong-04}
 and in real space (images recorded by atomic force 
 microscopy).~\cite{Thompson-08}
 Only lately, the attention has turned to the study of the domain structures
 in PbTiO$_3$/SrTiO$_3$ \emph{superlattices}.
 Zubko and coworkers~\cite{Zubko-10}
 have recently focused on 
 the dependence of their structural and dielectic properties as a 
 function of the volume fraction 
 of PbTiO$_{3}$, the electrodes, and the applied electric fields. 
 While the results of the superlattices asymmetrically sandwiched 
 between Nb-doped SrTiO$_{3}$ (bottom) and gold (top) 
 electrodes were consistent 
 with a monodomain configuration,~\cite{Dawber-05.2}
 those corresponding to the use of 
 symmetrically coated SrRuO$_{3}$ electrodes (both top and bottom) 
 suggested a polydomain phase with DW motion, which
 dynamics might be quite different than the conventional one.~\cite{Jo-11} 
 Furthermore, through a combination of x-ray diffraction, transmission electron
 microscopy, and ultra-high resolution
 electron energy loss spectroscopy (EELS), 
 Zubko {\it et al.}~\cite{Torres-Pardo-11,Zubko-12} 
 have also recently showed the progressive transition between two different
 regimes, controlled by the 
 thickness of the individual SrTiO$_{3}$ and PbTiO$_{3}$ layers.
 In the first regime, present for sufficiently thin paraelectric layers,
 SrTiO$_{3}$ and PbTiO$_{3}$ are strongly electrostatically \emph{coupled}:
 a uniform monodomain polarization is adopted throughout the thickness of 
 the superlattice to minimize the depolarizing field.
 In the second regime, when the paraelectric layer thickness is increased,
 SrTiO$_{3}$ and PbTiO$_{3}$ are \emph{decoupled}: 
 the polarization is confined within the
 FE PbTiO$_{3}$ layers forming domains. 
 EELS measurements revealed the presence of broad
 interfacial layers with reduced tetragonality and polarization extending over
 5-6 unit cells (u.c.) into the PbTiO$_{3}$ layers.
 Strikingly, in the electrostatic decoupled regime, the domain structure 
 was found to be coherent over several tens of superlattice repetitions.
 These works pointed out
 the suitability of this system for the study of 
 domains in ultra-thin ferroelectric films, given the behavior 
 of the ferroelectric layers 
 as quasi-independent films, while the thickness of the whole system
 prevents the charge leakage when electric fields are applied.~\cite{Zubko-10}
 They also constitute a good example of how to tune the functional properties
 with respect to different factors, such as the electrical
 boundary conditions or the periodicity of the superlattice. 

 On the other hand, the theoretical studies of 
 PbTiO$_{3}$/SrTiO$_{3}$ superlattices have 
 focused, so far, on three different aspects of the monodomain configurations:  
 (i) the analysis of the structural, electronic
 and ferroelectric properties of both 
 the pristine~\cite{Dawber-05.2,Cooper-07,Gu-10} 
 and disordered~\cite{Cooper-07,Gu-10}
 (including the presence of cation intermixing or defects) interfaces,
 (ii) the study of the coupling between the AFD instabilities and the FE 
 polarization, compatible with an improper ferroelectric 
 behaviour,~\cite{Bousquet-08} or 
 (iii) the influence of strain on the previous coupling.~\cite{Aguado-Puente-11}

 In this work we perform first-principles calculations on 
 polydomain phases of (PbTiO$_{3}$)$_n$/(SrTiO$_{3}$)$_n$ 
 [$(n\mid n)$] superlattices, 
 with periodicities of $n$ =3 and 6.
 Our goal is to gain further insight on the polarization 
 and oxygen octahedra rotation profiles.
 Besides, we compare the differences in energies between relevant phases.
 The influence of the periodicity, orientation, 
 energy of the DWs, and the mixed 
 FE-AFD-strain coupling 
 present in these superlattices~\cite{Aguado-Puente-11}  
 are carefully considered.
 We also analyze the strain fields induced by 
 the domain structure and their role in the inter-layer coupling.

 The rest of the paper is organized as follows:
 the method on which the simulations are based is decribed in 
 Sec.~\ref{sec:methods}.
 In Sec,~\ref{sec:energetics} we compare the energy of the 
 different competing phases (monodomain versus polydomain) to 
 ascertain their relative stability.
 The atomic structure of the domains is analyzed in 
 Sec.~\ref{sec:atomicstructure}.
 Finally, the polarization profiles and strain fields are discussed
 in Sec.~\ref{sec:polprofile}.

\section{Methods}
\label{sec:methods}

 The simulations have been carried out 
 within the local density approximation (LDA)
 to the density functional theory (DFT)
 using the {\sc siesta} code.~\cite{Soler-02}
 The rest of the technical parameters remain the same
 as in Ref.~\onlinecite{Aguado-Puente-11}. 

 In this work we have performed simulations of $(n\mid n)$ superlattices,
 by means of a supercell approach. Two values of $n$ have been 
 considered, $n=3$ and 6, aiming to sample superlattices within the 
 two distinct regimes experimentally observed:
 strong (for $n \lesssim 4$) and weak (for $n \gtrsim 4$) 
 electrostatic coupling between the SrTiO$_{3}$ and PbTiO$_{3}$ layers.

 As the starting point, an ideal structure was defined 
 stacking along the [001] direction $n$ unit cells of SrTiO$_{3}$ 
 and $n$ unit cells of PbTiO$_{3}$.
 The in-plane lattice constant was
 fixed to the theoretical LDA value of SrTiO$_{3}$ (3.874 \AA ).
 First, mirror symmetry planes were imposed at the central atomic layers
 of PbTiO$_{3}$ and SrTiO$_{3}$, and an initial atomic relaxation
 was performed in order to find a reference paraelectric ground-state. 
 Then, the reference structure was replicated $N_{x}$ times along the 
 [100] direction and $N_{y}$ times along the [010] direction. 
 Due to the periodic boundary conditions used in the simulations,
 $N_{x}$ determines the domain periodicity, while $N_{y}$ allows 
 to switch on ($N_{y}$ = 2) and off ($N_{y}$ = 1) the AFD instabilities. 
 Following the recipe given in Ref.~\onlinecite{Aguado-Puente-08}, a percentage
 of the bulk soft mode distortion 
 was superimposed on the PbTiO$_{3}$ layers, so the polarization points 
 upwards in half of the superlattice and downwards in the other half.
 For $N_y = 2$ superlattices, small rotations were induced by hand 
 following a $a^0a^0c^-$ pattern in Glazer notation.
 Finally, an extra atomic relaxation of the full heterostructure 
 was carried out,
 until the maximum value of the Hellman-Feynman forces
 and the $zz$ stress tensor component
 fell below 0.01 eV/{\AA} and 0.0001 eV/\AA$^3$ respectively
 [except for the $(3\mid 3)$ superlattice
 with $N_x=16$ and $N_y=2$ (960 atoms in the simulation box) 
 and for the $(6\mid 6)$ superlattice with $N_x=12$ and $N_y=1$ (720 atoms),
 which were relaxed down to a maximum force of 0.05 eV/{\AA}].

 To stablish the notation, we will call the plane parallel to the interface
 the $(x,y)$ plane, whereas the perpendicular direction will be referred to 
 as the $z$-axis. 

\section{Results}
\subsection{Energetics.}
\label{sec:energetics}

 \begin{figure}[]
    \begin{center}
       \includegraphics[width=\columnwidth]{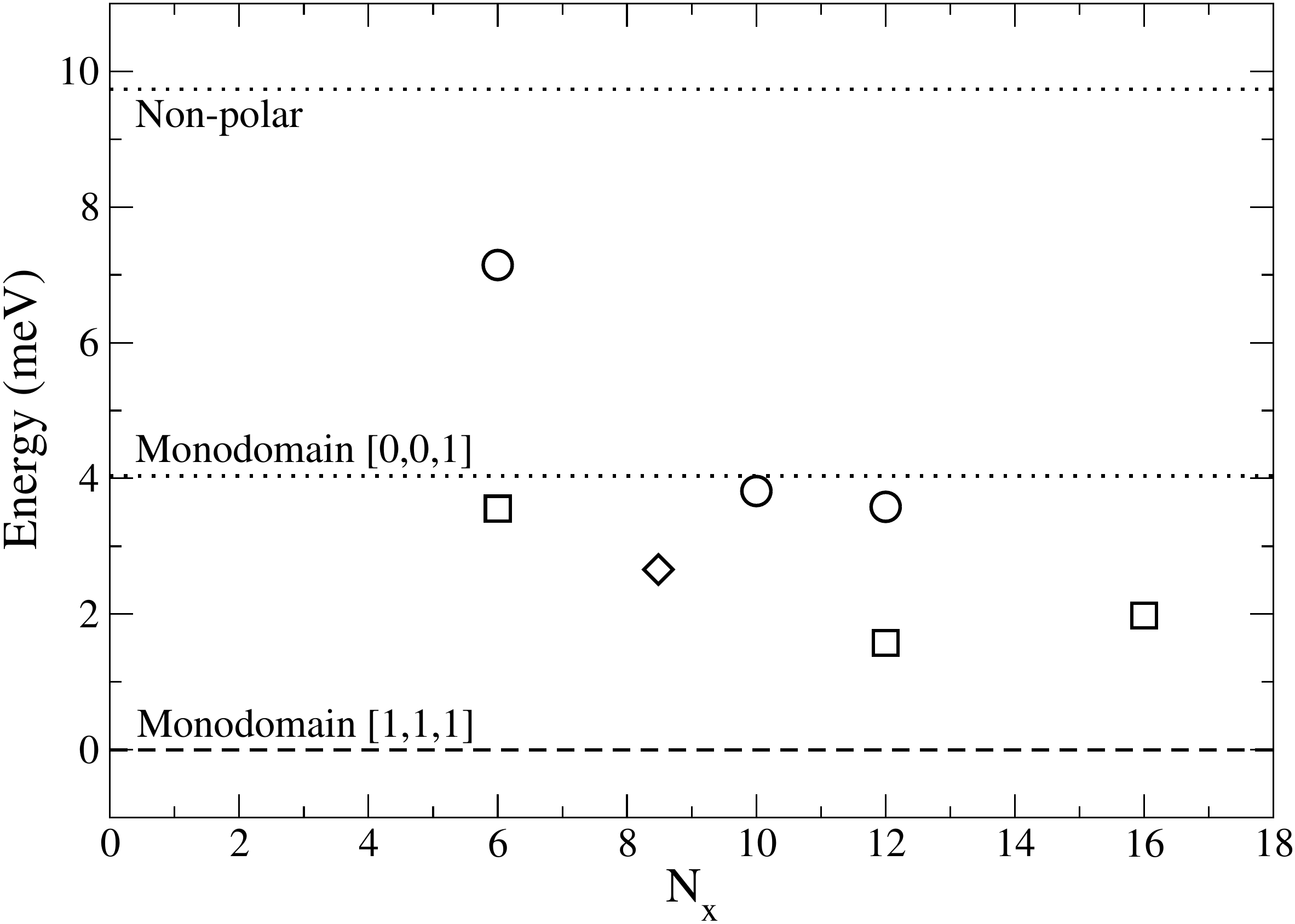}
       \caption{ Differences in energies between polydomain, monodomain and 
                 non-polar configurations in (3$\mid$3)
                 PbTiO$_3$/SrTiO$_3$ superlattices,
                 as a function of the domain period $N_{x}$. 
                 Total energies of supercells are given per 5-atom 
                 perovskite unit cell. 
                 Circles represent the configurations where the AFD modes are 
                 not allowed ($N_{y}$ = 1), 
                 while squares represent configuration with condensed
                 AFD modes ($N_{y}$ = 2).
                 Diamond indicates a configuration where the DW
                 lies along the $<110>$ direction, also allowing for the 
                 condensation of AFD modes. 
                 The monodomain phases have been labeled as in 
                 Ref.~\onlinecite{Aguado-Puente-11}.
                 In the non-polar configuration, the AFD distortions 
                 have been considered.
                 All energies are given with respect to the 
                 most stable monodomain configuration.
           }
       \label{fig:EvsNx}
    \end{center}
 \end{figure}

 For the $(3\mid 3)$ superlattices we have performed simulations of
 the different competing phases in order to determine their relative stability.
 The energies of the polydomain, monodomain and 
 non-polar configurations as a function of the domain periodicity 
 are shown in Fig.~\ref{fig:EvsNx}.
 For these superlattices,
 the balance between the electrostatic energy (which tends to reduce 
 the domain period), and the DW energy density 
 (which tends to increase it) results in an optimum periodicity 
 of the domain structure, $\Lambda$, of about 12 unit cells (46.5 \AA )
 (the energy for $N_{x}$ equal 12 and 16 might be considered as
 equivalent within the accuracy of our simulations).

 The most stable phase found in our simulations, however, corresponds to 
 a monodomain structure, with the polarization in the PbTiO$_{3}$
 layer pointing close to the perovskite unit cell diagonal 
 (configuration labeled as [111] in Ref.~\onlinecite{Aguado-Puente-11} and
 Fig.~\ref{fig:EvsNx}). 
 This result is consistent with the upturn in the domain periodicity
 observed by Zubko and coworkers,
 \cite{Zubko-12} suggesting that for $n<4$ the superlattices enter 
 into the strong coupling regime.
 Nevertheless, the energy difference
 between the monodomain and the most stable polydomain configuration
 is very small (of the order of 1.6 meV/5-atom-perovskite-unit-cell, well below
 the thermal energy at room temperature), suggesting a 
 close competition between them for small values of $n$.
 A small change on any external condition (growth temperature, 
 or how fast the system is cooled, etc.)
 might help the system to overcome potential energy barriers,
 activate transitions between them and could, eventually, stabilize 
 a metastable domain structure. This fact might explain 
 why both, polydomain and monodomain samples, have
 been observed experimentally.~\cite{Zubko-10}

\begin{figure}[]
    \begin{center}
       \includegraphics[width=\columnwidth]{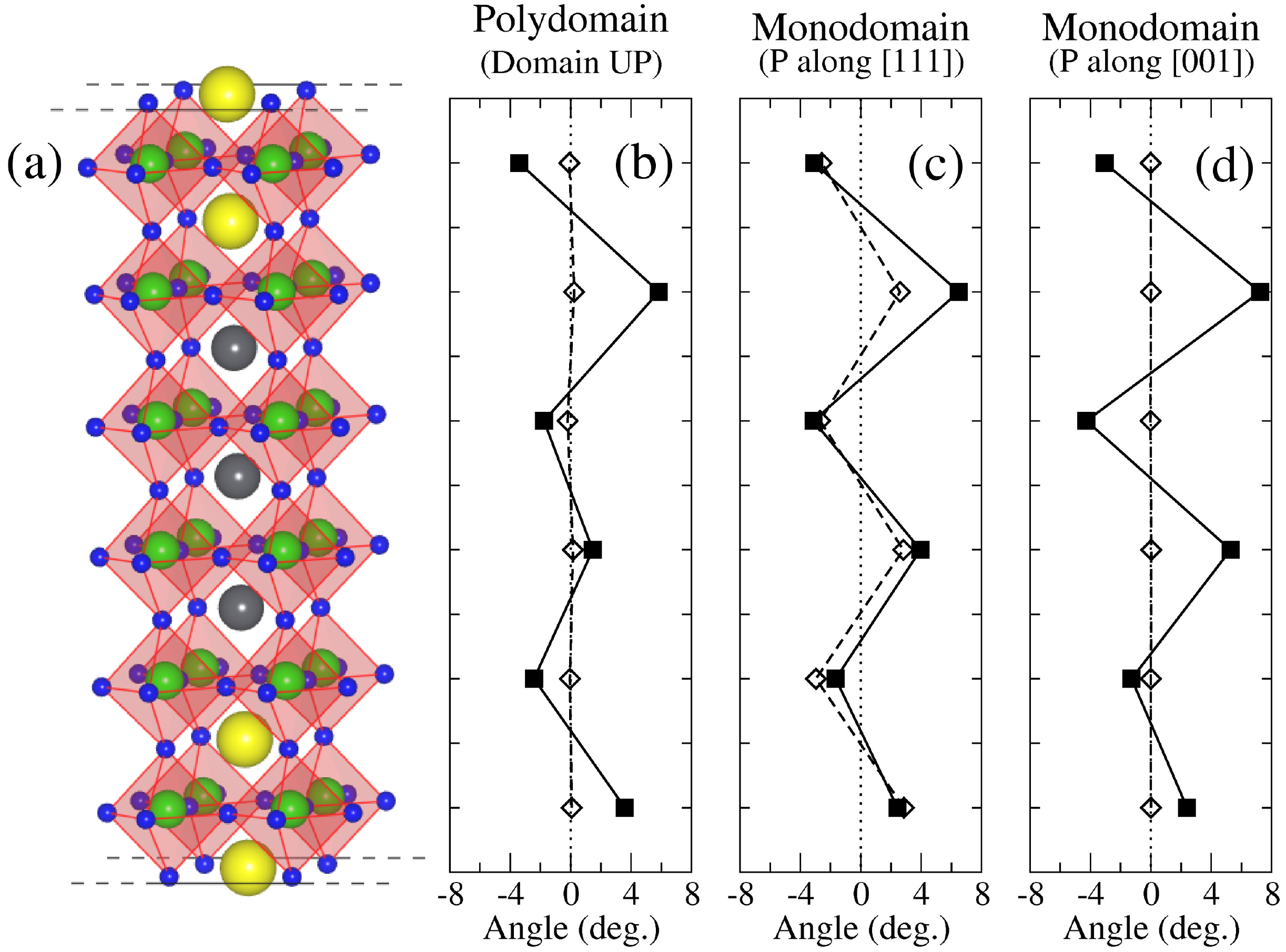}
       \caption{ (Color online) (a) Squematic representation of the 
                  center of a domain in a (3$\mid$3) PbTiO$_{3}$/SrTiO$_{3}$ 
                  superlattice (see region embodied by a bracket in 
                  Fig.~\ref{fig:esquema}). 
                  Atoms are represented by balls: Sr in yellow, Ti in green, 
                  O in blue, and Pb in grey. 
                  In panels (b)-(d) we represent the 
                  amplitude of the rotations (squares)
                  and tiltings (diamonds) of each TiO$_{6}$ 
                  octahedra: (b) at the center of a domain in the
		  polydomain configuration with $N_x=12$,
		  (c) in the ground state monodomain phase (with
		  polarization in the PbTiO$_{3}$ layer pointing close
                  to the perovskite unit cell diagonal, 
                  see Ref.~\onlinecite{Aguado-Puente-11}), and
                  (d) in a monodomain phase with polarization lying along
                  [001]. 
               }
       \label{fig:rotaciones}
    \end{center}
 \end{figure}

 For a given domain periodicity, the energy is systematically 
 lowered if the rotation of the oxygen octahedra are allowed, with reductions 
 ranging between 3.6 meV per 5-atom perovskite unit cell 
 (for $N_{x}$ = 6) to 2.0 meV (for $N_{x}$ = 12).  
 This highlights the importance of the FE-AFD coupling
 in these heterostructures.
 The coupling is also noticeable when the pattern of the oxygen 
 octahedra in the polydomain phases are compared with those of 
 monodomain configuration. We see in Fig. \ref{fig:rotaciones}
 that at the center of the domains (i. e. mid-distance
 between two DW, see column of atoms embodied by a bracket 
 in Fig.~\ref{fig:esquema}) where the polarization is
 purely out-of-plane, $P_{z}$, the rotations along an in-plane axis 
 (tiltings) essentially vanish, resembling the case of the [001] monodomain
 phase reported in Ref.~\onlinecite{Aguado-Puente-11}. Regarding the 
 rotations around the $z$-axis, 
 the larger $P_{z}$ in the PbTiO$_3$ layer in polydomain phases 
 penalizes the AFD modes and the rotation angles are smaller.
 The FE-AFD coupling also affects the magnitude of the polarization,
 resulting in a slight reduction of $P_z$ when condensation of
 AFD modes is allowed (see Table~\ref{tab:polyz}).

 \begin{table}[t]
    \caption{ Out-of-plane polarization, $P_{z}$, at the center of the domains
              in PbTiO$_{3}$/SrTiO$_{3}$ superlattices 
	      with $N_x=12$ u.c. 
              $P_{z}^{\rm PTO}$ ($P_{z}^{\rm STO}$) stands for the polarization
              at the central perovskite unit cell within the PbTiO$_{3}$ 
              (SrTiO$_{3}$) layer.
              Values in parenthesis correspond to
              the root mean square polarization, averaged along the 
              [100] direction. Units in $\mu$C/cm$^{2}$.
            }
    \begin{center}
      \begin{tabular*}{0.8\columnwidth}{@{\extracolsep{\fill}} ccccc}
         \hline \hline
         $(n\mid n)$                         &
         $N_{x}$                             &
         $N_{y}$                             &
         $P_{z}^{\rm PTO}$                   &
         $P_{z}^{\rm STO}$                   \\
         \hline
         $(3\mid 3)$ &
         12       &
         1        &
         65 (56)      &
         31 (26)      \\
         $(3\mid 3)$ &
         12       &
         2        &
         60 (53)      &
         29 (24)      \\
         $(6\mid 6)$ &
         12       &
         1        &
         75 (70)      &
         21 (17)      \\
         \hline
         \hline
      \end{tabular*}
   \end{center}
   \label{tab:polyz}
 \end{table}

 We also find that the effect of the DW orientation is small: a change 
 in the orientation of the DW from $<100>$ to $<110>$ does not 
 affect significantly the energy of the superlattice.
 This points to a rather isotropic DW structure,
 with the energy of the domains depending very weakly on the
 stripe orientation, in good agreement
 with experimental results,~\cite{Zubko-10}
 phenomenological Landau-Ginzburg-Devonshire theory,~\cite{Bratkovsky-11}
 and model Hamiltonian~\cite{Lai-07} simulations.
 
 For the $(6\mid 6)$ superlattices only one domain periodicity was
 simulated due to the scaling of the system size. For the same reason,
 in this case rotations of the oxygen octahedra were not allowed 
 in the calculations of the polydomain phases.
 In view of the results for the $(3\mid 3)$ superlattice, the presence
 of the octahedra rotations results in a decrease of the energies of 
 polydomains structures and a small reduction of the polarization 
 at the center of the domains. We chose $N_x = 12$ u.c.
 ($\Lambda = 46.5$ \AA), close to the experimental value 
 of $\Lambda = 55$ {\AA}.~\cite{Torres-Pardo-11}  
 These polydomain phase were
 found to lie 3.8 meV/5-atom-perovskite-unit-cell \emph{below}
 the most stable monodomain phase without 
 AFD distortions.

 The change in the most stable phase [from monodomain in the 
 $(3 \mid 3)$ to polydomain in the $(6\mid 6)$] indicates a crossover
 between the weak and strong electrostatic coupling regimes 
 described in the Introduction. Experimentally this transition
 was inferred to occur gradually, with a minimum of the domains size 
 observed at $n \simeq 4$ (Ref.~\onlinecite{Zubko-12}).
 In related KTaO$_{3}$/KNbO$_{3}$ superlattices, the critical periodicity
 for the crossover ranges between
 $ 7 \le n \le 15$ (experiments from Ref.~\onlinecite{Specht-98}), 
 and $12 \le n \le 24$ (shell models 
 simulations from Ref.~\onlinecite{Sepliarsky-01}).
 
\subsection{Atomic structure of the domains.}
\label{sec:atomicstructure}

 \begin{figure}[]
    \begin{center}
    \begin{overpic}[width=0.7\columnwidth]{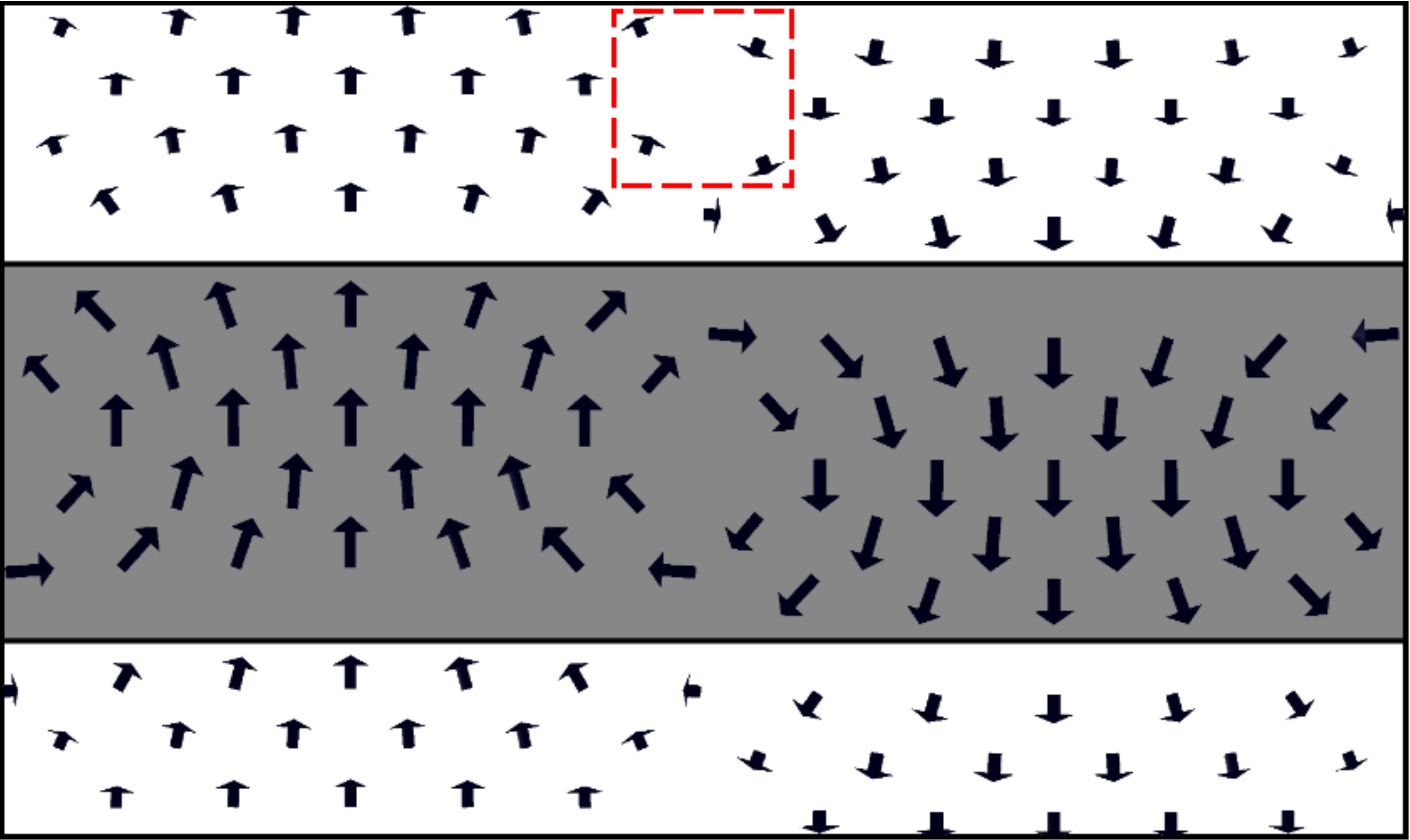}
      \put(-10,50){\large (a)}
    \end{overpic} \\
    \vspace{0.3cm}
    \begin{overpic}[width=0.7\columnwidth]{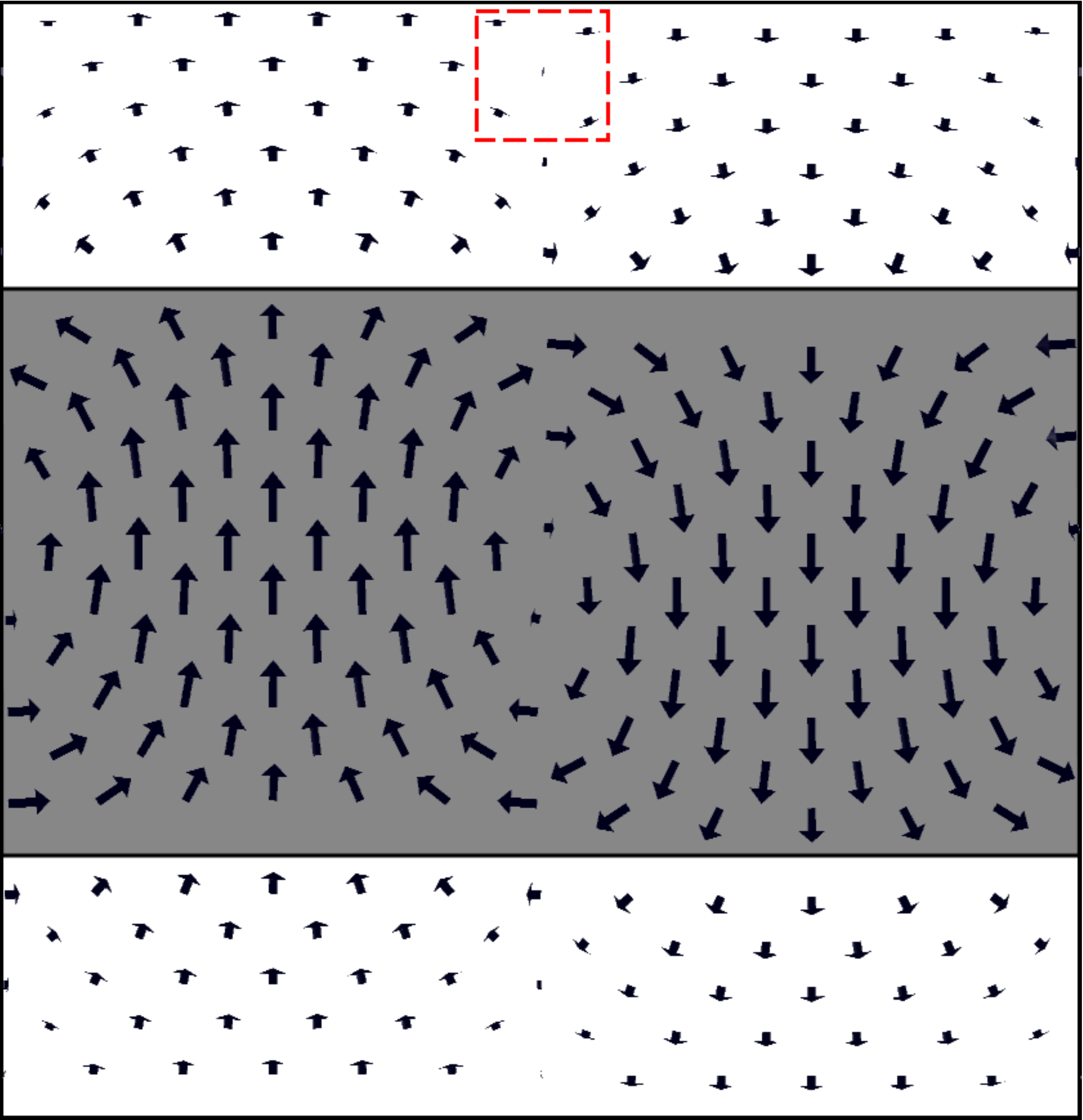}
      \put(-10,90){\large (b)}
    \end{overpic}
       \caption{ (Color online) Local polarization profile 
                 of polydomain structures in
                 (PbTiO$_3$)$_n$/(SrTiO$_3$)$_n$ superlattices with (a) $n=3$
                 and (b) $n=6$. 
                 The PbTiO$_3$ and SrTiO$_3$ are depicted as
		 gray and white regions respectively.
                 Red dashed squares in the SrTiO$_{3}$ layers mark the position
                 where antivortices are formed.}
       \label{fig:superlattice_vortices}
    \end{center}
 \end{figure}

 Figure \ref{fig:superlattice_vortices} shows the local polarization
 profile of
 the most stable polydomain configurations found for 
 the ($3\mid 3$) and ($6\mid 6$) superlattices (both with a 
 domain periodicity of $N_{x}$ = 12 u.c.). Similar patterns are
 obtained for other domain sizes. In order to
 be able to make quantitative comparisons, from now on only simulations
 where oxygen octahedra were \emph{not} allowed are discussed.
 The local polarization is obtained calculating the polarization of 
 a unit cell centered on every cation of the system (except
 at the interfaces where no ``bulk-like'' unit cell can be chosen)
 and using the displacement of the atoms with respect to the
 ideal phase and the bulk Born effective charges of the corresponding 
 material, either PbTiO$_3$ or SrTiO$_3$, depending on the
 layer the cation belongs to.

 Near the DWs the local polarization pattern clearly displays 
 a continuous polarization rotation
 within 3 u.c. around the DW, connecting two 180$^{\circ}$
 domains.

 As suggested by the experimental results exposed in Ref.~\onlinecite{Zubko-12},
 the examination of Fig. \ref{fig:superlattice_vortices} 
 reveals that the actual domain structure
 in this kind of systems is often an intermediate case between
 180$^\circ$ domains 
 and the closure domains commonly found in ferromagnets, displaying 
 rotation of the polarization upon approaching the DWs
 \cite{Kittel-46, Lee-09} (although the length scale over which the
 polarization rotation takes place is only a few unit cells in the
 case of ferroelectrics, in contrast with the several nanometers
 or even microns typical of ferromagnets).
 However, we have to keep in mind that in ideal closure domains,
 the divergence of the polarization vanishes everywhere and, therefore,
 the depolarizing field is perfectly screened.
 In our simulations, the polarization of the SrTiO$_{3}$ indicates
 the presence of a residual depolarizing field and thus,
 strictly speaking, our domains 
 do not constitute perfect domains of closure.

 Our results also support the robustness of the rotation of polarization and 
 the formation of vortices in 
 ferroelectric nanostructures suggested by previous
 theoretical studies. 
 These geometries have been predicted to exist independently
 of (i) the used methodology (including 
 first-principles,~\cite{Aguado-Puente-08,Shimada-10}
 model hamiltonians,~\cite{Lai-06, Prosandeev-07, Sichuga-11}
 phase field models,~\cite{Slutsker-08} and 
 phenomenological Devonshire-Ginzburg-Landau 
 theories~\cite{Stephenson-06,Bratkovsky-11}),
 and/or (ii) the electrostatic boundary conditions
 (with metallic~\cite{Aguado-Puente-08,Lichtensteiger-11} 
 or semiconducting electrodes,~\cite{Stephenson-06,Prosandeev-07} or even
 in free standing slabs.~\cite{Shimada-10, Sichuga-11})

 It is remarkable to see that the polarization rotation in the
 PbTiO$_3$/SrTiO$_3$ superlattices is mostly due to large
 in-plane displacement of the Pb atoms at the PbO layers
 in the vicinity of the interface. 
 This contrasts with the predicted domains in BaTiO$_{3}$/SrRuO$_{3}$
 capacitors,~\cite{Aguado-Puente-08} where the in-plane polarization
 is due to the displacements of the Sr atoms in the first layer of the 
 electrode.
 Here, the Pb atoms move of the order of 0.2 \AA , 
 a displacement large enough to be detectable with
 the recently developed atomic-resolution aberration-corrected
 transmission electron microscopy.
 Using this technique, polarization rotation at DWs have been
 experimentally observed 
 in ferroelectric thin films with thicknesses of a few
 tens of unit cells.~\cite{Nelson-11, Jia-11} However 
 the high quality level achieved during the last years in the growth of 
 short-period superlattices,
 together with the large in-plane displacements
 predicted, make this kind of system particularly
 suited for the observation of the formation of vortices at domain
 walls in ultrathin films, comparable in size to the simulated systems 
 listed above.

 Interestingly, within the SrTiO$_{3}$ layer and close to the DW
 we do observe the formation of antivortices; a local polarization
 pattern where two dipoles point face to face and two tail to tail 
 (see red dashed squares in Fig.~\ref{fig:superlattice_vortices}. 
 These antivortices
 have also been recently predicted to form in epitaxial BiFeO$_{3}$ 
 films.~\cite{Prosandeev-11})

\subsection{Polarization profiles and strain field}
\label{sec:polprofile}

 Within a polydomain configuration, there is no need anymore to keep constant 
 the normal component of the polarization at the interface, $P_{z}$, 
 since the electric fields that arise from its discontinuity are
 efficiently screened by the presence of domains. As a consequence $P_{z}$,
 that in the monodomain configuration is continuous throughout
 the superlattice, 
 in the polydomain case is very inhomogeneous with 
 polarization mismatches at the center of the domains of 
 34 $\mu$C/cm$^{2}$ and 54 $\mu$C/cm$^{2}$
 for the $(3\mid 3)$ and $(6\mid 6)$ superlattice respectively 
 (see Table~\ref{tab:polyz}). 
  
\begin{figure}[]
    \begin{center}    
    \begin{overpic}[height=\columnwidth,angle=270]{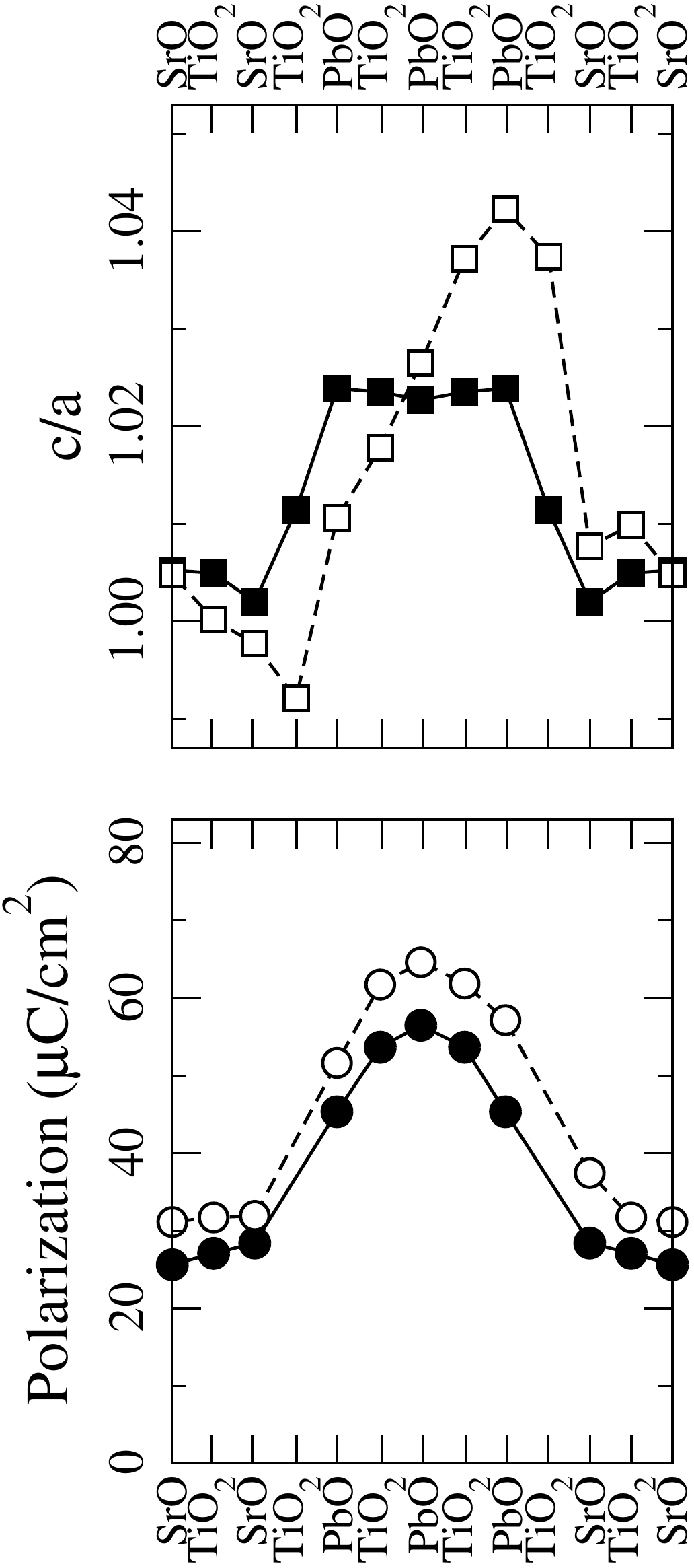}
      \put(38,7){\large (a)} \put(84.5,5.5){\large (b)}
    \end{overpic} \\
   \vspace{0.3cm}
    \begin{overpic}[height=\columnwidth,angle=270]{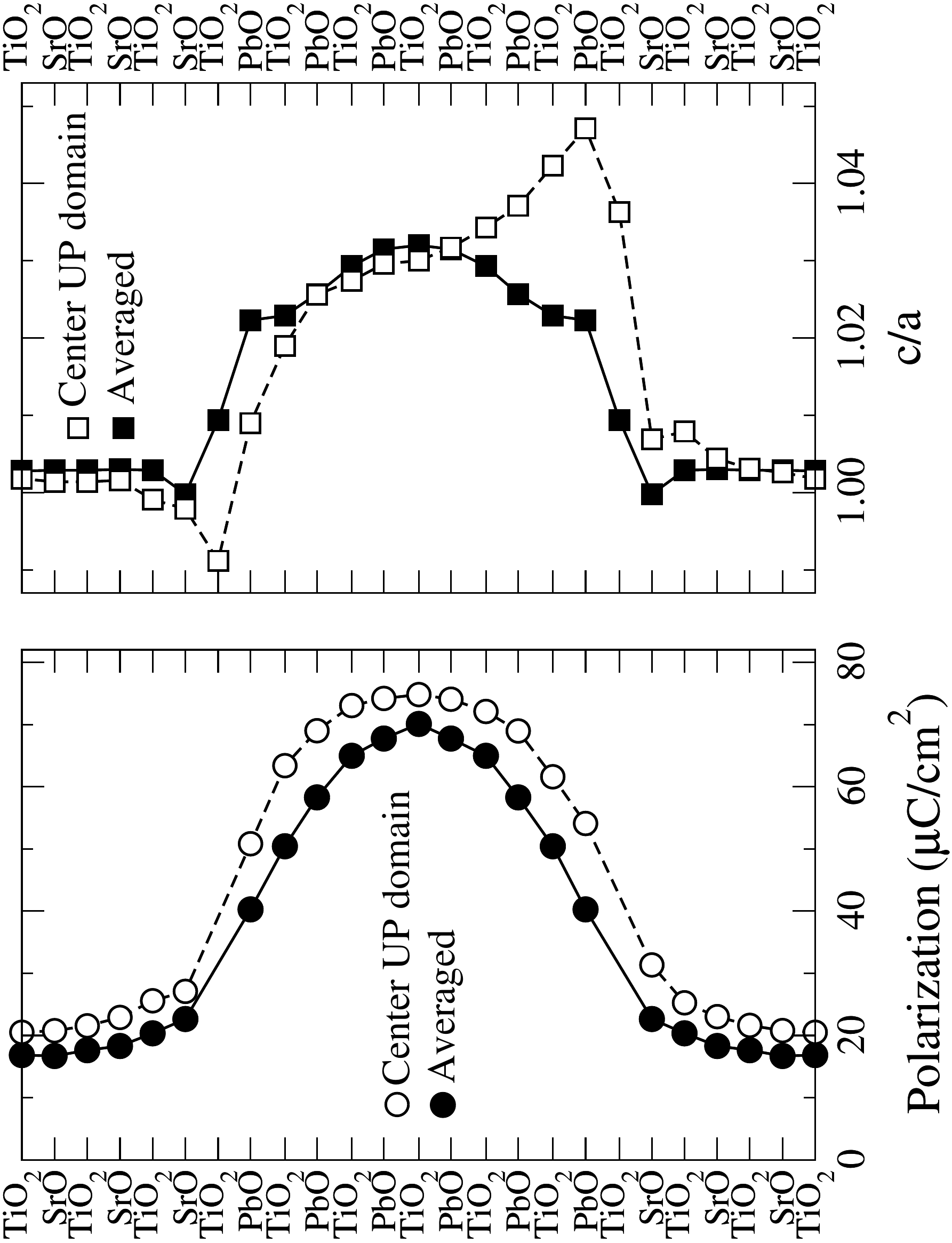}
      \put(38,17){\large (c)} \put(82,17){\large (d)}
    \end{overpic}
       \caption{  Left panels: layer-by-layer out-of-plane polarization,
                  $P_{z}$, inferred from the 
                  Born effective charges and the atomic displacements
                  for (a) a $(3\mid 3)$ and (c) a $(6\mid 6)$ superlattice. 
                  Right panels: layer-by-layer tetragonality for 
                  (b) a $(3\mid 3)$
                  and (d) a $(6\mid 6)$ superlattice.
                  Empty symbols represent values at the center 
                  of an up domain, while
                  filled symbols correspond to averaged values 
                  (root mean square in 
                  the case of polarization) along the [100] direction.
               }
       \label{fig:P_tetra}
    \end{center}
 \end{figure}

 The layer-by-layer polarization of Fig.~\ref{fig:P_tetra}(a) and 
 ~\ref{fig:P_tetra}(c)
 shows that, within SrTiO$_{3}$, it converges to a rather homogeneous
 well defined value, that at the center of the domain decreases from
 $31$ $\mu$C/cm$^{2}$ (for $n$ = 3) to $21$ $\mu$C/cm$^{2}$ 
 (for $n$ = 6). 
 On the contrary, the PbTiO$_{3}$ layer displays 
 a smooth variation of the polarization, with a progressive
 reduction spanning over a length of three unit cells into the PbTiO$_{3}$
 layers from the interface.
 The great reduction of the polarization of the SrTiO$_3$ layer upon
 a increase in $n$, together with the 
 out-of-plane polarization at the center 
 of the PbTiO$_3$ layer 
 rapidly approaching the bulk value (83 $\mu$C/cm$^{2}$),
 again supports the gradual
 electrostatic decoupling of the ferroelectric layer.~\cite{Zubko-12}

\begin{figure}[]
   \begin{overpic}[width=\columnwidth]{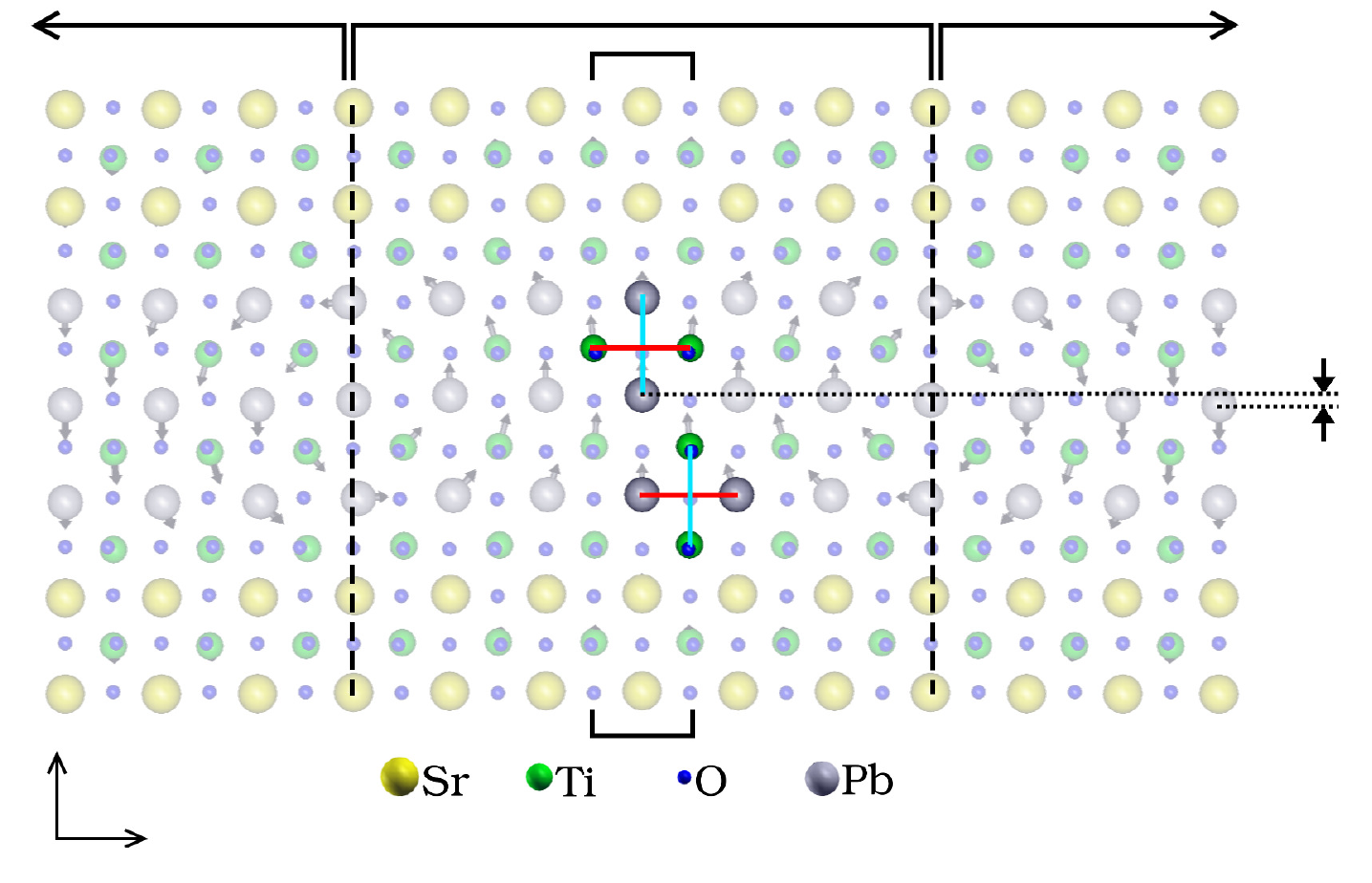}
     \put(44,25.5){$\boldsymbol{a_1}$}
     \put(51.5,32){$\boldsymbol{c_1}$}
     \put(40,36.5){$\boldsymbol{a_2}$}
     \put(48,43){$\boldsymbol{c_2}$}
     \put(90,29){offset}
     \put(42,65){\colorbox{white}{UP}}
     \put(7,65){\colorbox{white}{DOWN}}
     \put(72,65){\colorbox{white}{DOWN}}
     \put(4,-.5){[100]}
     \put(-1,4){\begin{sideways}[001]\end{sideways}}
   \end{overpic}
      \caption{(Color online) Schematic view of the $(3\mid 3)$ superlattice 
               with $N_x = 12$ and $N_y=1$ 
               indicating how local values of the magnitudes 
               plotted in Fig. \ref{fig:P_tetra}
               and \ref{fig:a_offset} are defined. 
               Red (blue) lines represent local values of in-plane, $a$, 
               (out-of-plane, $c$) lattice constants,
               measured from the in-plane (out-of-plane) 
               distance between equivalent cations
               of the same chemical specie in consecutive unit cells along
               the $x$ ($z$) direction.
               Magnitudes with subscript 1 (2) indicate unit cells centered on a
	       $[001]$ AO (TiO$_2$) atomic plane.
               Local polarization is marked with
               arrows.
               Black dotted lines indicate the offset between [100] atomic rows
               to the left and right of the domain walls,
               defined as the relative vertical shift of A-cations 
               in a given atomic plane.
               Bracket at the bottom
               of the up domain indicates the position 
               of its center, where the
               values plotted as empty symbols in Fig. \ref{fig:P_tetra} 
               and \ref{fig:a_offset} are obtained.		
               Finally, domain walls are represented by dashed lines. 
              }
      \label{fig:esquema}
\end{figure}
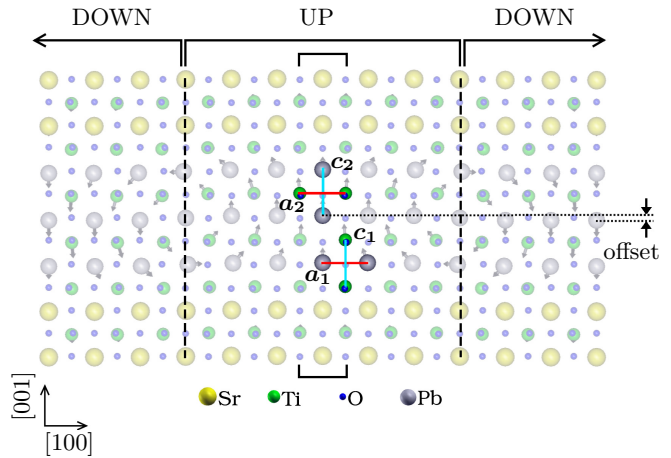

 In Fig. \ref{fig:P_tetra} we also plot
 the variation of the local tetragonality across the superlattice,
 calculated for the same perovskite unit cell surrounding
 each cation.  
 The layer-by-layer tetragonality \emph{averaged} along the [100] direction,
 plotted as black squares in 
 Fig. \ref{fig:P_tetra}(b) and (d), displays a variation that is well
 correlated with that of the polarization: an almost
 constant value inside the SrTiO$_3$ layer and a smooth
 increase from the interfaces towards the center of the PbTiO$_3$.
 The polarization of the SrTiO$_3$ layer induces a slight tetragonality 
 of this material. This reduction of symmetry with respect to
 the cubic unit cell of bulk SrTiO$_3$ is consistent with
 the decrease of the $t_{2g}$-$e_{g}$ splitting observed in
 the EELS spectra of this system by Zubko and coworkers.~\cite{Zubko-12}

 Besides, in Fig. \ref{fig:P_tetra}(b) and (d) we show
 the layer-by-layer tetragonality across the
 superlattice at the center of a domain with the polarization
 pointing \emph{up}.
 Strikingly, and contrary to the polarization, the variation of the local
 tetragonality as we move across the superlattice is very asymmetric. The
 tetragonality reaches its maximum value in the PbTiO$_3$ layer
 at the bottom interface with respect to the polarization direction,
 and gradually reduces its magnitude as we move towards the top interface.
 In the SrTiO$_3$ layer,
 the strain gradient is smaller and, forced by the 
 imposed periodic boundary conditions,
 it has opposite sign.

 The analysis of the polarization and tetragonality profiles 
 reveals that the formation of domains in the superlattices
 is associated with complex distortions. The characteristics of
 the strain field in this system can be explained as a
 combination of different effects.

\begin{figure}[]
    \begin{center}
       \includegraphics[width=\columnwidth]{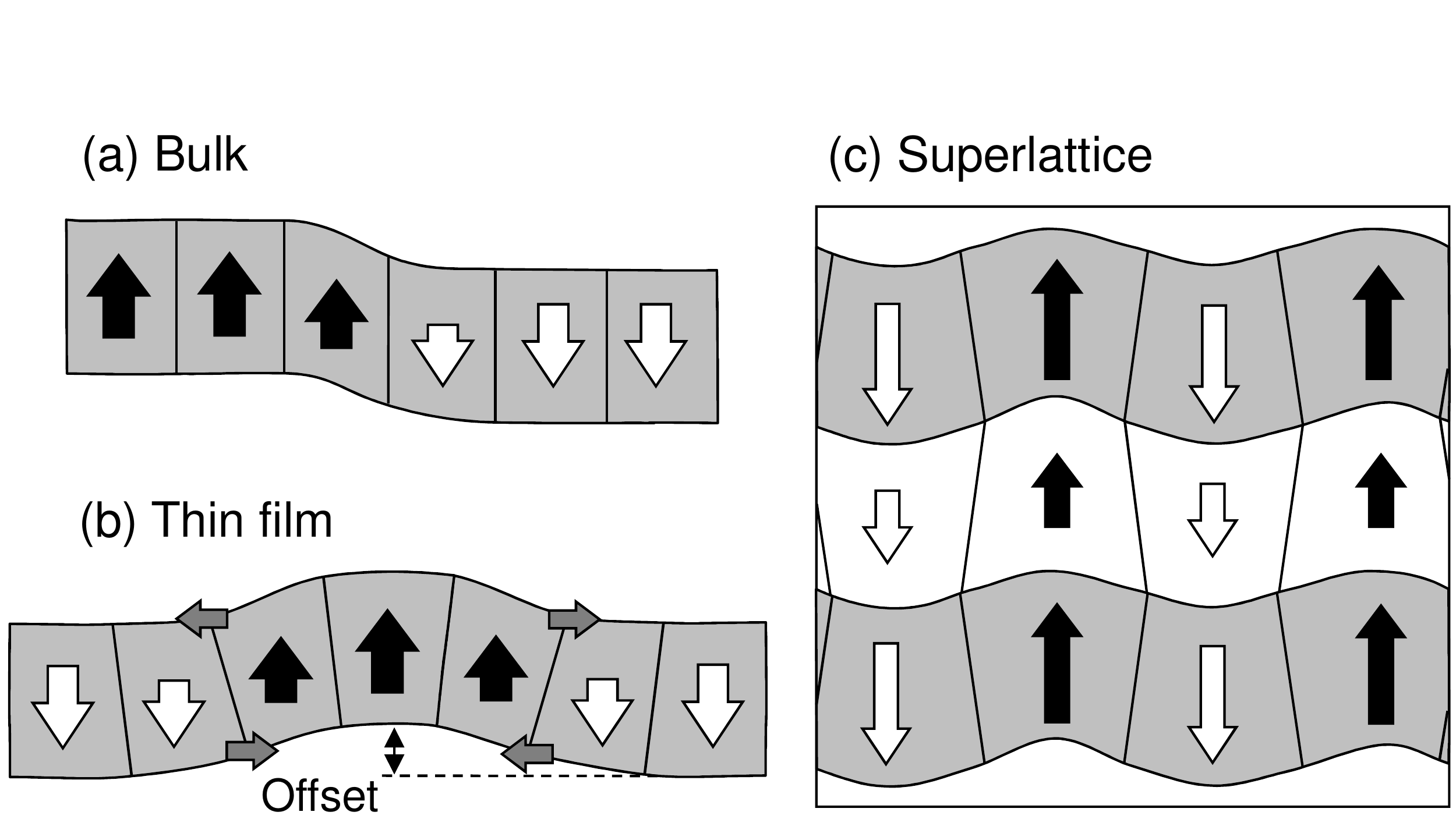}
       \caption{ Schematic representation of the distortion induced
                 by the domain structure in (a) bulk PbTiO$_3$,
                 (b) PbTiO$_3$ thin films and 
                 (c) PbTiO$_3$/SrTiO$_3$ superlattices. 
                 (a) In bulk, displacements of Pb cations cause an
                 offset between [100] atomic rows across the DW. 
                 (b) In thin films, in addition to the offset between domains,
                 rotation of the polarization near the interface is responsible
                 of a non-vanishing strain gradient
                 $\frac{\partial \varepsilon_{11}}{\partial z}$.
                 (c) In the case of the PbTiO$_3$/SrTiO$_3$ 
                 superlattices, the offset and modulation of
                 the strain field in the PbTiO$_3$ layer 
                 (in grey) propagates 
                 into the SrTiO$_3$ (in white).
               }
       \label{fig:wiggle}
    \end{center}
 \end{figure}

\begin{figure}[]
    \begin{center}
    \begin{overpic}[height=\columnwidth,angle=270]{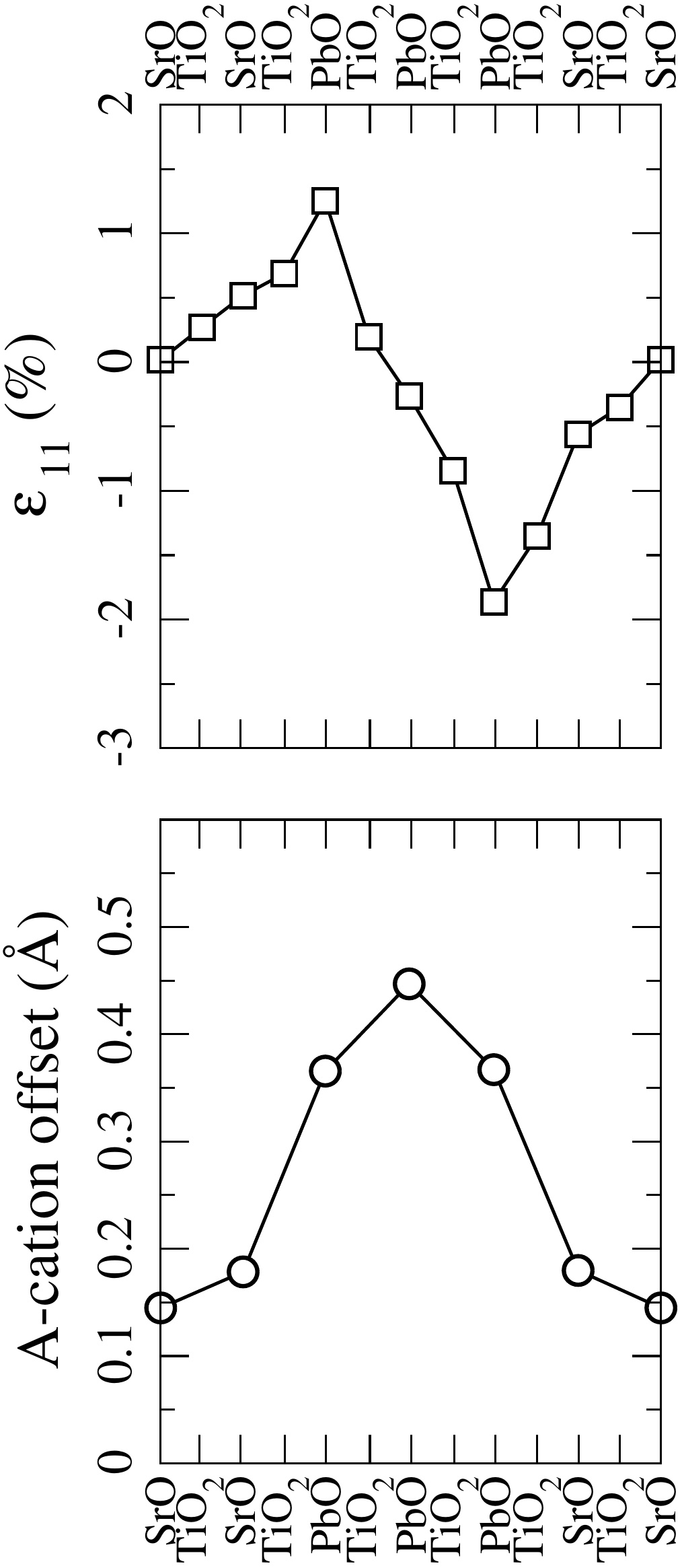}
      \put(38,7){\large (a)} \put(82,7){\large (b)}
    \end{overpic} \\
   \vspace{0.3cm}
    \begin{overpic}[height=\columnwidth,angle=270]{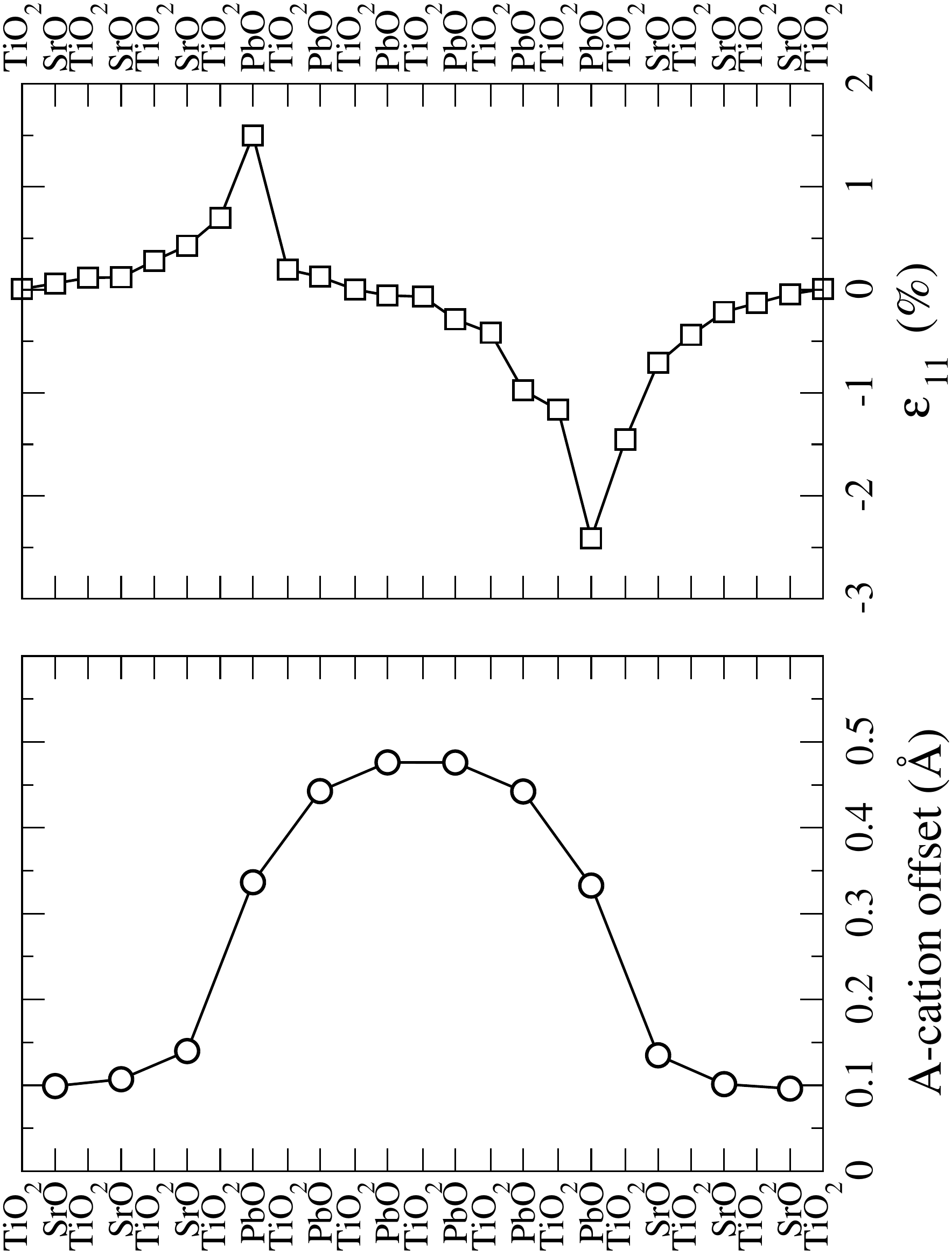}
      \put(38,17){\large (c)} \put(82,17){\large (d)}
    \end{overpic}
       \caption{  Left panels: local layer-by-layer offset 
                  between $[100]$ atomic rows to the left and right 
                  of the DW for (a) a $(3\mid 3)$ and 
                  (c) a $(6\mid 6)$ superlattice.
                  Right panels: local in-plane strain  
                  across the center of an up domain in (b) a $(3\mid 3)$
                  and (d) a $(6\mid 6)$ superlattice. A large non-diagonal
                  component of the strain gradiend, 
                  $\frac{\partial \varepsilon_{11}}{\partial z}$ can be
                  observed close to the interfaces. 
               }
       \label{fig:a_offset}
    \end{center}
 \end{figure}

 On the one hand, in PbTiO$_3$ the off-center displacements of 
 both the Pb and Ti cations contribute to the polarization. Therefore
 the Pb atoms displace along $z$ in opposite direction in the up and
 down domains. This gives rise to an \emph{offset} between
 $[100]$ atomic rows to the left and right of the DW
 [see Fig. \ref{fig:wiggle}(a)]. A sizable offset of 0.6 \AA\
 was already predicted by Meyer and Vanderbilt in 180$^\circ$ 
 stripe domains in bulk PbTiO$_3$.~\cite{Meyer-02} 
 As in Ref. \onlinecite{Meyer-02}, for the PbTiO$_3$/SrTiO$_3$ superlattices 
 we quantify this offset for a given layer as the difference in the 
 $z$-coordinate of a equivalent A-cation at the center of opposite domains
 (see Fig.~\ref{fig:esquema}). 
 The layer by layer offset, shown in
 Fig. \ref{fig:a_offset}(a) and (c), amounts up to almost 0.5 (0.45) {\AA} 
 at the middle of the 
 PbTiO$_3$ layer in the $(6\mid 6)$ $[(3\mid 3)]$ superlattice.
 Although the offset of opposite domains 
 is partially accommodated by the interfaces -- which reflects in
 the increase (decrease) of the tetragonality at the bottom (top)
 interface in Fig. \ref{fig:P_tetra}(b) and (d) --, it still propagates
 into the SrTiO$_3$, amounting a sizable $\sim 0.1$ {\AA}.

 On the other hand, in thin films the polarization in
 PbTiO$_3$ rotates at the DW. Indeed, as pointed out above,
 our simulations show that large in-plane displacements
 of the Pb atoms (up to 0.2 \AA) take place a the interfaces.
 It is sensible to argue that this in-plane polarization
 is coupled with an in-plane strain and, as it is
 schematically depicted in Fig. \ref{fig:wiggle}(b),
 it pushes the DW in the same direction of the polarization. 
 This effect is reinforced as consecutive DWs become closer,
 as it happens in ferroelectric thin films [Fig. \ref{fig:wiggle}(b)].
 As a consequence, the in-plane lattice constant is 
 expected to be enlarged
 at the top interface (with respect to polarization direction)
 and compressed at the bottom interface.
 To test this hypothesis 
 we have performed a detailed analysis of the strain field in
 the system, calculating for every individual
 perovskite unit cell the local
 values of the in-plane lattice constant, $a$ (see Fig.~\ref{fig:esquema} for
 an explanation about how it is computed). 
 The local in-plane strain, calculated as 
 $\varepsilon_{11}= a/a_{0}-1$, where $a_{0} = a_{\rm SrTiO_{3}} = 
 3.874$ \AA\ is plotted
 in Fig. \ref{fig:a_offset}(b) and (d)
 for the $(3\mid 3)$ and $(6\mid 6)$ superlattices respectively.
 It shows a variation with respect to the position along 
 the $z$-direction
 that can be clearly correlated with that of the tetragonality, shown in
 Fig.~\ref{fig:P_tetra}(b) and (d):
 PbTiO$_3$ unit cells close to the bottom interface
 (with respect to the polarization direction) are compressed
 in-plane, and as a result, 
 they tend to elongate along the $z$ axis.
 Conversely, at the top interface, the 
 material is expanded in-plane and
 presents a reduced tetragonality. 
 Since the in-plane polarization is confined at the
 interfaces, large strain gradients 
 $\frac{\partial \varepsilon_{11}}{\partial z}$
 can be anticipated. In fact, huge values 
 are obtained from our simulations: up to $4\cdot 10^7$
 m$^{-1}$, more than seven orders 
 of magnitude larger than those obtained in bending 
 experiments on SrTiO$_3$~\cite{Zubko-07}.
 Similar distortions are found in the SrTiO$_3$ layer,
 although with opposite sign of the strain gradients.

 Finally, in the superlattice, the SrTiO$_3$ atomic layers
 closer to the interface are forced to follow the local in plane
 expansion or contraction of the PbTiO$_3$ [see Fig. \ref{fig:wiggle}(c)].
 However, the presence of an inhomogeneous strain in the SrTiO$_3$ is 
 energetically very costly. Therefore it rapidly recovers a nearly constant 
 in-plane lattice parameter [see Fig. \ref{fig:a_offset}(d)].

 The combined effect of the domain offset and the modulation
 of the strain field can explain the
 large out-of-plane coherence length of domain structure 
 previously observed in this superlattice.~\cite{Zubko-12} Even for 
 relatively large thicknesses of SrTiO$_3$,
 for which the ferroeletric layers can be considered
 as being electrostatically decoupled, domains with the same orientation of 
 the polarization are aligned along the [001] direction. 

 The large strain gradients associated with
 the polydomain configurations might,
 in addition, significantly affect the local polarization pattern through the 
 flexoelectric effect.~\cite{Cross-06,Hong-10,Catalan-11,Lee-11,Hong-11} 
 According to the sign
 of the gradients relative to the polarization direction
 the flexoelectric effect tends to increase the polarization
 in the PbTiO$_3$ and decrease it in the SrTiO$_3$.~\cite{Bursian-68,Bursian-04}
 A numerical quantification of the corresponding non-diagonal 
 component of the flexoelectric tensor is extremely subtle, 
 since it might be hidden by strain contributions
 via piezoelectric effects, and is out of the scope of this paper.
 However, assuming a flexoelectric coefficient of
 the order of 1 nC/m (typical
 of ferroelectric perovskites \cite{Zubko-07}), and with the
 strain gradients extracted above, we estimate that the 
 flexoelectric-induced polarization can reach values of 
 a few $\mu$C/cm$^2$ at the interfaces or at the center of 
 the short period superlattices.

\section{Conclusions}

 In summary, 
 using accurate first-principles simulations we have
 studied the domain structures in short-period 
 (PbTiO$_{3}$)$_{n}$/(SrTiO$_{3}$)$_{n}$ superlattices.
 The most important conclusions that can be drawn are:
 (i) the domain structures might compete in energy with 
 monodomain configurations.
 (ii) The domains are rather isotropic, a change
 in the orientation of the DW from $<100>$ to $<110>$ does not
 affect significantly the energy of the superlattice.
 (iii) From the structural point of view, they display 
 polarization-rotation, similar to the one
 theoretically predicted in ferroelectric nanocapacitors and recently
 observed in various ferroelectric ultrathin films.
 (iv) Our results suggest the progressive transition as a function of $n$ 
 from a 
 strongly electrostatic coupled regime (where the ground-state 
 is a monodomain configuration
 with a constant out-of-plane component of the polarization preserved
 throughout the structure), to a weakly coupled regime 
 (where the polarization is confined within the PbTiO$_{3}$ layers forming
 domains). 
 (v) The evolution of the out-of-plane layer-by-layer 
 polarization and tetragonality within the
 SrTiO$_{3}$ and the PbTiO$_{3}$ layers
 are consistent with the
 $t_{2g}$-$e_{g}$ splitting inferred from the unit-cell-resolution
 recently measured by EELS experiments.~\cite{Zubko-12}
 (vi) Large offsets between [100] atomic rows across the DW
 and huge strain gradients (seven orders of magnitude larger than those
 obtained in bending experiments on SrTiO$_3$~\cite{Zubko-07}) are observed.
 The contribution of both of them can be responsible of the out-of-plane
 coherence of the domain structure found experimentally.~\cite{Zubko-12}

 This knowledge should complement the experimental studies 
 and could encourage the design of new artificial structures with even more
 appealing functionalities.

 The authors thank Dr. P. Zubko for the valuable discussion and
 his careful reading of the manuscript.
 This work was supported by the Spanish Ministery of Science and
 Innovation through the MICINN Grant FIS2009-12721-C04-02, by the
 Spanish Ministry of Education through the FPU fellowship AP2006-02958 (PAP),
 and by the European Union through the project EC-FP7,
 Grant No. CP-FP 228989-2 ``OxIDes''.
 The authors thankfully acknowledge the computer resources, 
 technical expertise and assistance provided by the 
 Red Espa\~nola de Supercomputaci\'on.
 Calculations were also performed at the ATC group
 of the University of Cantabria.

\end{document}